\documentclass[nofootinbib, reprint]{revtex4} 
\usepackage{amssymb, amsmath}
\usepackage{graphicx}
\usepackage{verbatim}
\begin{document}
\title{The effect of time-dilation on Bell experiments in the retrocausal Brans model}
\author{Indrajit Sen}
\email{isen@g.clemson.edu}
\affiliation{Department of Physics and Astronomy,
Clemson University, Kinard Laboratory,
Clemson, SC 29634, USA}
\date{\today}

\begin{abstract}
The possibility of using retrocausality to obtain a fundamentally relativistic account of the Bell correlations has gained increasing recognition in recent years. It is not known, however, the extent to which these models can make use of their relativistic properties to account for relativistic effects on entangled systems. We consider here a hypothetical relativistic Bell experiment, where one of the wings experiences time-dilation effects. We show that the retrocausal Brans model (Found. Phys, 49(2), 2019) can be easily generalised to analyse this experiment, and that it predicts less separation of eigenpackets in the wing experiencing the time-dilation. This causes the particle distribution patterns on the photographic plates to differ between the wings -- an experimentally testable prediction of the model. We discuss the difficulties faced by other hidden variable models in describing this experiment, and their natural resolution in our model due to its relativistic properties. In particular, we discuss how a $\psi$-epistemic interpretation may resolve several difficulties encountered in relativistic generalisations of de Broglie-Bohm theory and objective state reduction models. Lastly, we argue that it is not clear at present, due to technical difficulties, if our prediction is reproduced by quantum field theory. We conclude that if it is, then the retrocausal Brans model predicts the same result with great simplicity in comparison. If not, the model can be experimentally tested.
\end{abstract}
\maketitle

\section{Introduction}
While Einstein could not find a satisfactory realist account of quantum theory in his lifetime, he managed to deliver a fatal blow to the complacency surrounding the theory's interpretation by his EPR argument \cite{epr, einstein48, honstein}, which gave rise in turn to Bell's theorem \cite{bell}. The theorem proves under certain assumptions that any realist interpretation of quantum mechanics must be nonlocal\footnote{This does not mean that the theorem disproves `local realism', which is a misleading expression. See ref. \cite{norse} for a discussion.}. One key assumption of the theorem is that the hidden variable distribution satisfies the `measurement-independence' \cite{hall10, howmuch, hall16} condition
\begin{align}
\rho(\lambda|\psi, M) = \rho(\lambda|\psi) \label{moo}
\end{align}
where the hidden variables are labelled by $\lambda$, the preparation by the quantum state $\psi$, and the measurement setting by $M$. The assumption states that the hidden variable distribution has no correlation with the future measurement settings. This assumption has often gone unnoticed in the vast literature surrounding Bell's theorem, but in the recent past has come under increasing scrutiny. Foremost among the possible implications is violating the assumption to build a realist account of quantum theory which satisfies locality, thereby fulfilling Einstein's dream \cite{schilp}.\\

There are at least two physical mechanisms to violate measurement-independence. The first is superdeterminism, a proposal mentioned by Bell \cite{gosht}. In a superdeterministic model, the hidden variable distribution and the measurement settings are statistically correlated due to past initial conditions. Bell criticised these models as `conspiratorial' \cite{dialect}, and they have since been widely criticised as such in the literature, barring a few exceptions \cite{brans, hautomaton}. We give a comprehensive overview of the various arguments related to superdeterministic models, and quantify the notion of the alleged conspiracy, in a forthcoming paper \cite{forthy}. \\

The other physical option is retrocausality \cite{costacoffee, cramer, pricebook, pricebash, sutherland, whartonmain}; that is, a model where the future measurement setting causally affects the hidden variable state in the past. Highly counter-intuitive to our ordinary sense of time and causality, such models nevertheless provide perhaps our best bet to build a realist description of quantum theory which is fundamentally relativistic. A recent survey \cite{whartreview} has pointed out the infancy of the retrocausal project: only a limited number of quantum phenomena have been explicitly described by such models yet. One possible option to cover more ground is to model relativistic phenomena in entangled systems. Retrocausal models are especially well-suited to this task as they are already relativistic at the hidden variable level. This will also answer a question of utility: can retrocausal models \textit{utilise} their relativistic properties? In this article, we answer the question in affirmative for a particular scenario involving the relativistic effect of time-dilation on a Bell experiment. We show that the retrocausal Brans model \cite{fpaper} can be easily generalised to describe this scenario, and that it predicts an experimentally testable consequence of the time-dilation effect on the system. We also discuss the difficulties faced by other hidden variable models in describing this experiment, and their natural resolution in our model due to its relativistic properties: locality, separability and $\psi$-epistemicity \cite{harrikens, leifer, pbr}. In particular, our model suggests that several problems in proposals for relativistic generalisations of de Broglie-Bohm theory \cite{bohm1, bohm2, solventini} and objective state reduction models \cite{grw, pearl, grp} may be resolved by a $\psi$-epistemic interpretation. Lastly, we argue that a description of this experiment in quantum field theory is a much more technically challenging task than in the retrocausal Brans model. \\

The article is structured as follows. In section \ref{anna} we describe our scenario and point out why non-relativistic quantum mechanics cannot be used to describe it. We briefly recapitulate the retrocausal Brans model in section \ref{manna}. In section \ref{khanna}, we generalise this model and apply it to the present scenario. We show that the model predicts a particular difference in the position distributions between the wings due to the time-dilation effect. In section \ref{nanha}, we discuss the difficulties faced by other hidden variable models in describing our scenario, especially focusing on $\psi$-ontic multi-time models \cite{bell22, hyper96, l7, tumulkastic}, as they bear a structural similarity to the model we present in section \ref{khanna}. We discuss our results in section \ref{ranna}. We discuss which properties of our model are instrumental in resolution of the difficulties encountered in section \ref{nanha}, the implications for relativistic generalisations of hidden variable models, and the technical difficulties in describing this experiment using quantum field theory. We end with a comment on some of the important challenges facing the retrocausal approach, including the problem of fine-tuning \cite{woodspek}. 

\section{A relativistic Bell experiment}\label{anna}
\begin{figure}
\includegraphics[scale=0.5]{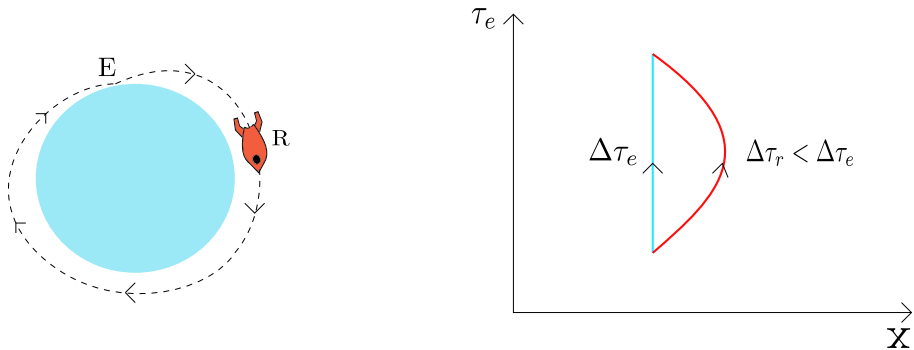}
\caption{Schematic illustration of the experiment. The first wing (E) of the experiment is located in a laboratory on Earth. The second wing (R) is located inside a relativistic rocket which is initially grounded close to the laboratory. A Bell experiment is now conducted, where the Stern-Gerlach magnets in both the laboratories are turned on simultaneously as the rocket lifts off into orbit. The magnetic fields are switched off in both the laboratories simultaneously when the rocket returns to the same spot. In general, the proper time elapsed in E, say labelled by $\Delta \tau_e$, is more than that elapsed in R, say labelled by $\Delta \tau_r$. How does this affect the experimental outcomes?}
\end{figure}

Consider a Bell experiment where the first wing is located in a laboratory on Earth, whereas the second wing is loaded onto a rocket which can move at relativistic speeds. Both the wings are equipped with Stern-Gerlach magnets to perform local spin measurements on their respective particles. The rocket is initially grounded on Earth adjacent to the laboratory. Suppose the spin measurements are started (that is, the Stern Gerlach magnetic fields are turned on) on both the laboratory and the rocket simultaneously at the instant the rocket lifts off into orbit. Let the rocket return to the same spot after a certain time, at which instant the magnetic fields are turned off in both the laboratory and the rocket (see Fig. 1). The question is: will there be any difference in the experimental results between the wings? We assume that all experimental parameters are such that the notion of a particle can still be approximately maintained in the relativistic domain. In other words, that relativistic phenomena like Unruh effect, spin-flip transitions, Wigner rotation of spin, radiation due to acceleration of charged particles etc., do not significantly affect the system. In particular, the acceleration of the rocket is assumed to be sufficiently gentle. This is not a fundamental limitation, however, as one can also consider an alternative version of the experiment which removes the play of acceleration altogether by using a cylindrical space-time (see Fig. 2).\\

\begin{figure}
\includegraphics[scale=0.5]{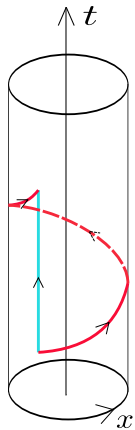}
\caption{Schematic illustration of an alternative version of the experiment in a cylindrical flat space-time. The position axis is wrapped around the surface of the cylinder, the time axis is vertical. A Bell experiment is conducted on two entangled spin-1/2 particles which travel along different world-lines (depicted in blue and red) that start off, and end, at common endpoints. Although neither of the particles are accelerated, the proper time durations elapsed along the world-lines are different due to the cylindrical geometry of space-time.}
\end{figure}

Let us attempt to describe the experiment using non-relativistic quantum mechanics. Say the spin measurement $\hat{\sigma}_r$ is conducted inside the rocket, and that $\hat{\sigma}_e$\footnote{$\hat{\sigma}_r \equiv \hat{\sigma} \cdot \hat{r}$ and $\hat{\sigma}_e \equiv \hat{\sigma} \cdot \hat{e}$} is conducted in the laboratory on Earth. Consider a von-Neumann measurement, where the interaction Hamiltonian\footnote{One may also consider the Stern-Gerlach interaction Hamiltonian $\hat{H}_I = \mu(B_o + B_1 \hat{z})\hat{\sigma}_z$. Since this does not lead to any additional physical insight, but makes the calculations longer, we stick to the von-Neumann interaction Hamiltonian. See section \ref{dhanna} for more details.} is of the form $\hat{H}_I =g\hat{p}\otimes\hat{\sigma}_z$. Here $g$ is a constant proportional to the strength of interaction. The total Hamiltonian of the system in non-relativistic quantum mechanics will then be
\begin{align}
\hat{H} = \big ( \frac{\hat{p}_r^2}{2m}\otimes \hat{I}\otimes \hat{I}\otimes \hat{I} + g\hat{p}_r\otimes \hat{\sigma}_r\otimes \hat{I}\otimes \hat{I}\big ) + \big (\hat{I}\otimes \hat{I}\otimes\frac{\hat{p}_e^2}{2m}\otimes \hat{I} + g\hat{I}\otimes \hat{I}\otimes\hat{p}_e\otimes \hat{\sigma}_e \big ) \label{aleph}
\end{align}

However, we know that this form of the Hamiltonian is an approximation valid only when the momenta $p_e$ and $p_r$ are very small compared to $mc$, where $c$ is the speed of light in vacuum. This is not possible in any \textit{single} frame of reference in which we choose to compute the total Hamiltonian. But we are restricted to a single frame as the total system is described by a \textit{joint} quantum state. Therefore, non-relativistic quantum mechanics cannot be applied to this experiment. However, we will see that the retrocausal Brans model, which gives a local and separable explanation of the Bell correlations in the non-relativistic limit, can be easily generalised to describe this experiment. In the next section, we briefly recapitulate this model.

\section{The retrocausal Brans model}\label{manna}
The model begins by positing that the information about the \textit{future} measurement settings, say labelled by $\hat{\sigma}_r$ and $\hat{\sigma}_e$, is made available to the particles at the preparation source in the \textit{past}, by an as yet not understood `retrocausal mechanism'. This causes the particles to be prepared in one of the eigenstates of the future measurement settings. That is, the pairs of particles are prepared in one of these joint spin states: $|+\rangle_r \otimes|+\rangle _e$, $|+\rangle_r \otimes |-\rangle _e$, $|-\rangle_r\otimes |+\rangle _e$, $|-\rangle_r \otimes|-\rangle _e$, where $|i\rangle_{r(e)}$ is an eigenstate of $\hat{\sigma}_{r(e)}$, $i \in \{+,-\}$.\\

Each particle is described in the model by an \textit{ontic quantum state} of the form $\chi(\vec{x},t)|i\rangle$, where $\chi(\vec{x},t)$ is a single particle 3-space wavepacket and $|i\rangle$ is an eigenstate of the future measurement setting. The pair of particles is described by the initial joint ontic quantum state $\langle \vec{x}_r| \langle \vec{x}_e|\psi_o(0)\rangle = \chi_r(\vec{x}_r,0)|i_1\rangle_r \otimes\chi_e(\vec{x}_e,0)|i_2\rangle_e$. We term the preparation-determined quantum state $\langle \vec{x}_r| \langle \vec{x}_e|\psi_e(0)\rangle=\chi_r(\vec{x}_r,0)\chi_e(\vec{x}_e,0)|\psi_{singlet}\rangle$ as the \textit{epistemic quantum state}. Both the joint ontic quantum state $\big ($with two single particle 3-space wavepackets $\chi_r(\vec{x}_r,t)$ and $\chi_e(\vec{x}_e,t)\big)$ and the epistemic quantum state $\big($with a single configuration space wavepacket $\chi_{re}(\vec{x}_r,\vec{x}_e, t)$ in general$\big)$ evolve via the multi-particle Schrodinger equation in the model. \\

The model next posits that each particle has a definite position at all times, with velocity given by a de Broglie-Bohm type guidance equation
\begin{align}
\vec{v} = \frac{\vec{\nabla}S(\vec{x},t)}{m}\label{d4}
\end{align}
where $\chi(\vec{x},t)=R(\vec{x},t)e^{iS(\vec{x},t)}$ is the 3-space wavepacket of that particle, contained in the ontic quantum state. The trajectory of the particle (and hence the measurement outcome) is thus determined locally by the single-particle ontic quantum state. This completes description of the ontology of the model.\\

For an ensemble of particles, we assume that the expansion of the preparation-determined (epistemic) quantum state in the future measurement basis 
\begin{align}
|\psi_{singlet}\rangle = c_{++}|+\rangle _r|+\rangle_e + c_{+-}|+\rangle_r|-\rangle _e
+ c_{-+}|+\rangle_r |+\rangle _e + c_{--} |-\rangle _r|+\rangle_e \label{d3}
\end{align}
determines the ensemble-proportions $|c_{++}|^2$, $|c_{+-}|^2$, $|c_{-+}|^2$, $|c_{--}|^2$ of the initial joint ontic quantum states $\chi_r(\vec{x}_r,0)|+\rangle_r\otimes \chi_e(\vec{x}_e,0) |+\rangle _e, \textbf{ }\chi_r(\vec{x}_r,0)|+\rangle_r\otimes \chi_e(\vec{x}_e,0)|-\rangle _e, \text{ }\chi_r(\vec{x}_r,0)|-\rangle_r \otimes \chi_e(\vec{x}_e,0) |+\rangle _e, \text{ }\chi_r(\vec{x}_r,0)|-\rangle_r \otimes \chi_e(\vec{x}_e,0) |-\rangle _e$ respectively. It can be shown that these proportions remain constant in time \cite{fpaper}. Secondly, we assume that the distribution of positions over the ensemble is given by
\begin{align}
\rho(\vec{x}_r,\vec{x}_e,t) = \sum_{i_1,i_2} |c_{i_1i_2}|^2 \rho_{i_1i_2}(\vec{x}_r,\vec{x}_e,t) \label{d5}
\end{align}
where $\rho_{i_1i_2}(\vec{x}_r,\vec{x}_e,0)=|\chi_r(\vec{x}_r,0)|^2|\chi_e(\vec{x}_e,0)|^2$ $\forall \textbf{ } i_1, i_2$ and $\rho_{i_1i_2}(\vec{x}_r,\vec{x}_e,t)$ is determined by the continuity equation implied by the time evolution of the joint ontic quantum state $\chi_r(\vec{x}_r,0) |i_1\rangle_r\otimes\chi_e(\vec{x}_e,0)|i_2\rangle_e$ via the multi-particle Schrodinger equation. Given these assumptions, the model correctly reproduces the Bell correlations \cite{fpaper}. In the next section, we apply this model to the relativistic case considered in this paper by an appropriate generalisation.

\section{Description of the experiment in the model}\label{khanna}
\subsection{Generalisation of the model to the relativistic case}
The key features of the model which allow for a relativistic generalisation are \textit{separability} and \textit{locality}. Each particle has a separate hidden variable of the form $(\vec{x}(t), \chi(\vec{x},t)|i\rangle)$. The position $\vec{x}(t)$ evolves locally via equation (\ref{d4}). The joint ontic quantum state of both the particles $\chi_r(\vec{x}_r,t)|i_1\rangle_r\otimes \chi_e(\vec{x}_e,t) |i_2\rangle _e$ evolves via the multi-particle Schrodinger equation. Since the non-relativistic Hamiltonian $\big ($equation (\ref{aleph})$\big )$ is separable, the individual ontic quantum states evolve locally. \\

It has been assumed that the time parameter $t$ is common to both the particles. However, this is an approximation valid only at the non-relativistic limit. In the relativistic case, we \textit{replace the common time parameter $t$ by the proper times at the wings}. That is, the hidden variable state of the particle in the rocket will be $(\vec{x}(\tau_r), \chi(\vec{x},\tau_r)|i_1\rangle_r)$, and that of the particle in the Earth laboratory will be $(\vec{x}(\tau_e), \chi(\vec{x},\tau_e)|i_2\rangle_e)$. Since the time parameters are different in both the wings, it is clear that the total non-relativistic Hamiltonian, given by eqn. (\ref{aleph}), cannot be used to evolve the joint ontic quantum state of the particles $\chi(\vec{x}_r,\tau_r)|i_1\rangle_r \otimes \chi(\vec{x}_e,\tau_e)|i_2\rangle_e$. However, we note that the total Hamiltonian is separable, and is of the form $(\frac{\hat{p}^2}{2m}\otimes \hat{I} +g\hat{p}\otimes \hat{\sigma})$ at each wing. This form of the local Hamiltonian at each wing is applicable even in the relativistic case since the relative speed of the particle with respect to the Stern-Gerlach apparatus remains small at each wing (though the speed of the other particle is large with respect to that wing). Therefore, it is natural to posit that the ontic quantum state of each particle evolves via a \textit{single-particle} Schrodinger equation which contains the local Hamiltonian describing that particular wing. That is, 
\begin{align}
\big ( \frac{\hat{p}_r^2}{2m}\otimes \hat{I} + g\hat{p}_r\otimes \hat{\sigma}_r \big ) \chi_r(\vec{x}_r,\tau_r)|i_1\rangle_r  = i\frac{\partial \chi_r(\vec{x}_r,\tau_r)|i_1\rangle_r}{\partial \tau_r} \label{01} \\
\big ( \frac{\hat{p}_e^2}{2m}\otimes \hat{I} + g\hat{p}_e\otimes \hat{\sigma}_e \big ) \chi_e(\vec{x}_e,\tau_e)|i_2\rangle_e  = i\frac{\partial \chi_e(\vec{x}_e,\tau_e)|i_2\rangle_e}{\partial \tau_e} \label{02}
\end{align}

Similarly, we modify the guidance equation (\ref{d4}) to
\begin{align}
\vec{v} = \frac{\vec{\nabla} S(\vec{x}, \tau)}{m} \label{aam}
\end{align}
where $\chi(\vec{x}, \tau) = R(\vec{x}, \tau) e^{iS(\vec{x}, \tau)}$ is the 3-space wavepacket of the particle, contained in the ontic quantum state\footnote{Strictly speaking, equations (\ref{01}) and (\ref{02}) imply continuity equations of the form $\frac{\partial R^2}{\partial t} + \vec{\nabla}\cdot\big ( R^2(\frac{\vec{\nabla}S}{m} + g \hat{s})\big ) =0$, where $\hat{s}$ is along the direction of spin-measurement. This implies that the exact guidance equation should be of the form $\vec{v} = \frac{\vec{\nabla}S}{m} + g \hat{s}$ for this particular Hamiltonian. We ignore this inessential detail for the purposes of this article.}.\\

In the non-relativistic case, the ensemble proportions of the ontic quantum states at any time $t$ were given by decomposition of the prepared quantum state at that time in the future measurement basis. For the relativistic case, we generalise the usual quantum state to a multi-time quantum state $|\psi(\tau_e, \tau_r)\rangle$ parametrised by the proper times at both the wings. Multi-time quantum states were originally introduced in the early days of relativistic quantum mechanics \cite{dirac32, bloch34, diracpofuck}. They were also discussed by Bell \cite{bell22} in the context of the GRW objective state reduction model, and continue to be discussed by multiple authors \cite{hyper96, hyper99, t7, l7}, especially in proposals for a fundamentally relativistic version of the de Broglie-Bohm theory. For our model, we posit, analogous to the non-relativistic case, that the decomposition of $|\psi(\tau_e, \tau_r)\rangle$ in the future measurement basis at any particular pair of proper times $(\tau_e, \tau_r)$ determines the ensemble proportions of the ontic quantum states at that $(\tau_e, \tau_r)$. Therefore, the role of the multi-time quantum state in our model is at all times purely statistical. We posit that the evolution of this multi-time quantum state is given by
\begin{align}
\big ( \frac{\hat{p}_r^2}{2m}\otimes \hat{I}\otimes \hat{I}\otimes \hat{I} + g\hat{p}_r\otimes \hat{\sigma}_r\otimes \hat{I}\otimes \hat{I} \big ) |\psi(\tau_e, \tau_r)\rangle  = i\frac{\partial|\psi(\tau_e, \tau_r)\rangle}{\partial \tau_r} \label{01'} \\
\big ( \hat{I}\otimes \hat{I}\otimes \frac{\hat{p}_e^2}{2m}\otimes \hat{I} + \hat{I}\otimes \hat{I}\otimes g\hat{p}_e\otimes \hat{\sigma}_e \big ) |\psi(\tau_e, \tau_r)\rangle  = i\frac{\partial |\psi(\tau_e, \tau_r)\rangle}{\partial \tau_e} \label{02'}
\end{align}
Note that both the equations refer to different reference frames. We also assume, as before, that the initial position distribution for an ensemble of particles on the rocket (Earth laboratory) is given by $|\chi_r(\vec{x}_r, 0)|^2 \big (|\chi_e(\vec{x}_e, 0)|^2\big )$. This distribution evolves via the continuity equation implied by the Schrodinger evolution of the ontic quantum state via equation (\ref{01}) $\big ($(\ref{02})$\big )$. The model can now be applied to the experiment.
\begin{figure}
\includegraphics[scale=0.25]{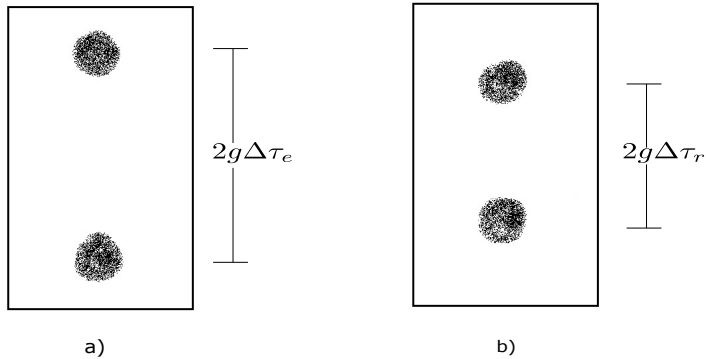}
\caption{Schematic illustration of the difference in position distributions between the wings. Part a) shows the photographic plate in the Earth laboratory, and part b) shows the photographic plate in the rocket after the measurements are over. Here $\Delta \tau_e$ and $\Delta \tau_r$ are the proper times elapsed on the Earth laboratory and the rocket respectively between the lift-off and return of the rocket.}
\end{figure}
\subsection{Description of the experiment}\label{dhanna}
The starting point of our investigation is the pair of equations (\ref{01}) and (\ref{02}). Let the constant $g$, which determines the strength of interaction at both the wings, be large enough such that the kinetic energy term $\frac{p^2}{2m}$ is negligible in comparison. In that case, the equations reduce to 
\begin{align}
 g\hat{p}_r\otimes \hat{\sigma}_r \chi_r(\vec{x}_r,\tau_r)|i_1\rangle_r  = i\frac{\partial \chi_r(\vec{x}_r,\tau_r)|i_1\rangle_r}{\partial \tau_r}  \\
g\hat{p}_e\otimes \hat{\sigma}_e \chi_e(\vec{x}_e,\tau_e)|i_2\rangle_e  = i\frac{\partial \chi_e(\vec{x}_e,\tau_e)|i_2\rangle_e}{\partial \tau_e} 
\end{align}

These equations can be solved to give
\begin{align}
\chi_r(\vec{x}_r,\tau_r)|i_1\rangle_r = \chi_r(\vec{x}_r - i_1 g\tau_r \hat{r}, 0)|i_1\rangle_r \\
\chi_e(\vec{x}_e,\tau_e)|i_2\rangle_e = \chi_e(\vec{x}_e - i_2 g\tau_e \hat{e}, 0)|i_2\rangle_e
\end{align}
Thus, similar to the non-relativistic case, the wavepackets separate in physical space depending upon the ontic spin states. The only difference is that the \textit{rate of separation is different at both the wings}, as it depends on the proper time elapsed. Since the proper time elapsed on the rocket is lesser than that on the Earth laboratory, the separation of the wavepackets will be lesser on the rocket. This will reflect as difference in the particle distribution patterns on the photographic plates between the wings (see Fig. 3).\\

The probabilities of the various outcomes remain to be ascertained. From equations (\ref{01'}) and (\ref{02'}), the multi-time epistemic quantum state evolves to
\begin{align}
\langle \vec{x}_r |\langle \vec{x}_e |\psi(\tau_r, \tau_e)\rangle &= \langle \vec{x}_r |\langle \vec{x}_e |e^{-ig\hat{p}_r\otimes \hat{\sigma}_r \otimes \hat{I}\otimes \hat{I} \tau_r} e^{-ig \hat{I}\otimes \hat{I}\otimes \hat{p}_e\otimes \hat{\sigma}_e \tau_e}|\psi(0, 0)\rangle\\
&=\sum_{i_1,i_2} c_{i_1, i_2}\chi_r(\vec{x}_r - i_1 g\tau_r \hat{r}, 0) \chi_e(\vec{x}_e - i_2 g\tau_e \hat{e}, 0) |i_1\rangle_r |i_2\rangle_e \label{multime}
\end{align}
Thus, we see that the proportions of ontic quantum states remain constant, and equal to $|c_{i_1, i_2}|^2$, for all $(\tau_r, \tau_e)$. The Bell correlations are therefore trivially reproduced, except in the extreme case that the wavepackets on the rocket do not sufficiently separate and therefore overlap each other, since then the measurement will have failed to be of sufficient resolution on the rocket. The shape of the spot on the photographic plate corresponding to the outcome $|i_1\rangle_r$ ($|i_2\rangle_e$) on the rocket (Earth laboratory) will be given by the distribution $|\chi_r(\vec{x}_r - i_1 g\tau_r \hat{r}, 0)|^2$ ($|\chi_e(\vec{x}_e - i_2 g\tau_e \hat{e}, 0)|^2$). \\

It might be pointed out that we have considered a von-Neumann measurement, which consists of an impulse-like interaction (lasting for a very short period of time) between the quantum system and the measuring apparatus. On the other hand, we assumed that this interaction lasts for the whole duration of the rocket trip, which is inconsistent with the assumption of an impulsive measurement. We have used the simplicity of von-Neumann measurement to bring out the central argument of the article without going into unnecessary complications. A more realistic version of the experiment, which is consistent with impulsive measurements, can be considered by using the Stern-Gerlach Hamiltonian, which is of the form $\hat{H}_I = \mu(B_o + B_1 \hat{z})\hat{\sigma}_z$. For this Hamiltonian, the wavepackets begin separating \textit{after} the magnetic field is turned off. Thus, one can modify the experiment such that the magnetic fields are turned on, for a very short duration, in both the wings while the rocket is grounded on Earth. Further, let the distance between the particle source and the photographic plate in both the wings be arranged such that the particles impact the photographic plates when the rocket completes the trip\footnote{This can be done because the motion of the particle through the Stern-Gerlach apparatus in a direction perpendicular to the magnetic field can be treated classically \cite{bohmbook}.}. After the magnetic fields are turned off, the rocket lifts off as usual. The spread of wavepackets (and the longitudinal distance travelled by the particle) will be less on the rocket during the trip due to the time-dilation effect, which will result in different position distributions on the photographic plates, as before. \\

\section{Description in other hidden variable models} \label{nanha}
To understand the physical significance of the prediction made in the previous section, we must consider modelling our scenario in other hidden variable models. A nonlocal model with a preferred reference frame (defining a single $t$ variable) will, clearly, face difficulties as the non-relativistic form of the Hamiltonians in equation (\ref{aleph}) cannot be valid in any single frame. This does not mean that nonlocal models defined on a preferred frame cannot, in principle, describe this experiment  \cite{bell9, largevale}, but that any adequate description will probably be much more complicated than the one presented in this article. A more interesting case to consider is a hidden variable model which (like our model) uses a multi-time quantum state $|\psi(\tau_r,\tau_e)\rangle$, but treats it as an ontological variable \cite{harrikens, pbr}. Prima facie, it might appear that the introduction of two time variables (one corresponding to each wing) in a hidden variable model is alone sufficient to account for the time-dilation effect. In this section, we argue that this impression is misleading. \\

Consider a simple model in which $|\psi(\tau_r,\tau_e)\rangle$, with its evolution given by equations (\ref{01'}) and (\ref{02'}), comprises the sole ontology. For this model, generalising the collapse postulate of orthodox quantum mechanics, according to which a local measurement collapses the entangled quantum state instantaneously across the frame in which the quantum state is defined, is problematic. This is because of two reasons. Firstly, the evolution of $|\psi(\tau_r,\tau_e)\rangle$ is defined on \textit{two} different reference frames via equations  (\ref{01'}) and (\ref{02'}). It is not clear in which of these frames (or any other) the collapse should be defined. Secondly, as the multi-time quantum state is ontological in this model, the collapse is a nonlocal process whereby a local measurement at one wing nonlocally affects the particle at the other wing \cite{harrikens}. That is, the particle at the other wing jumps from a mixed state before the collapse to a pure state after the collapse. Since the quantum state is ontological in the model, this is a real, physical transition. Thus, any frame in which the collapse may be defined would constitute a preferred frame of reference where this effect is instantaneous in the joint system, despite the cosmetic appearance of a multi-time quantum state defined on two different reference frames. But we know that both equations (\ref{01'}) and (\ref{02'}) cannot be true in any single reference frame. Thus, giving a clear and consistent formulation of this model in our scenario turns out to be problematic. \\

The above remarks also apply to multi-time formulation \cite{bell22, tumulkastic, t7} of objective state reduction models \cite{grw, pearl, grp} as these models feature an ontological quantum state which suffers a stochastically determined instantaneous collapse. An objective state reduction model with `flash' ontology \cite{tumulkastic}, however, can be reformulated so as to avoid instantaneous collapse of the ontic quantum state \cite{alluri}. It has not been possible, as of yet, to describe interactions in such a model, and consequently, the model fails to generate testable experimental predictions. This is because the relevant flashes occur in the model on practicable timescales only after an interaction between the quantum system and a macroscopic measuring apparatus. Still, it is of interest to ask if such a reformulated objective state reduction model can adequately describe our scenario if interaction is successfully incorporated in future. Let us, therefore, consider our scenario in this model after such an interaction has occurred, described by a joint entangled quantum state of the singlet pair and the photographic plates at both the wings. The probability of a flash occurring at a space-time point $(x, t)$ for one of the particles can then be calculated in the model by considering the linearly evolved multi-time quantum state on a space-time hypersurface containing the point $(x, t)$ and the past (with respect to the selected hypersurface) flashes. In general, this probability turns out to depend on the specific hypersurface considered. This is problematic for a subtle but important reason. There is a single relative frequency of flashes that actually occurs at $(x, t)$ for an ensemble. Although practically it is not possible, as of yet, to empirically determine this relative frequency, it exists in principle. The natural way to conceptually reconcile this single relative frequency with the multiple probabilities assigned by the model is to interpret these probabilities as representing different subjective information in different hypersurfaces about the flash process at $(x, t)$. But the flash probabilities, on the other hand, are supposed to arise from a fundamentally indeterministic process described by the ontic quantum state. The probabilistic description does not arise out of a subjective description about this process (because the quantum state is ontic), but from the fundamental indeterminacy of the process itself. That is, the flash probability at $(x, t)$ is supposed to give an \textit{objective} description of this fundamentally indeterministic process occurring at $(x, t)$. There should, therefore, be a one-to-one correspondence between this probability at $(x, t)$ and the space-time point $(x, t)$. But the model, as we discussed, does not predict any such unique quantity in our scenario. Thus, similar to the previous model, giving a clear and consistent formulation of the reformulated objective state reduction model is problematic for our scenario, regardless of whether interaction is included in the future. We note that this problem does not occur in non-relativistic objective state reduction models because they are formulated on a preferred hyperplane in space-time. \\

Lastly, let us consider adding particle positions to the ontology to resolve the problems associated with instantaneous collapse. In this model, the multi-time quantum state $\psi(\vec{x}_r, \tau_r; \vec{x}_e, \tau_e)$ acts as a nonlocal guiding wave for the particle positions. To define the nonlocal velocity field associated with this guiding wave, a preferred foliation of space-time is necessary. Several models \cite{bohmbook2, hyper99} of this kind have been proposed in the literature\footnote{There are also models \cite{hyper96, nikol, golduka, horton} featuring the multi-time quantum state as a guiding wave without any distinguished foliation. These models fail to predict any probabilities for experiments however (for a discussion of these issues, see ref. \cite{t7, durr14}). Therefore they cannot reproduce for our scenario the distribution of positions on the photographic plates, neither the difference between these distributions across the two wings. Consequently, we do not discuss them here.}. A reasonable foliation for our scenario would be one which is locally tangent to the simultaneity surfaces at each point of the rocket's trajectory as well as the Earth laboratory's trajectory (the foliation, therefore, has to be curved). On such a foliation, if it exists, the non-relativistic form of the local Hamiltonians corresponding to the different wings in equations (\ref{01'}) and (\ref{02'}) will be justified. This model will correctly reproduce the time-dilation effect on the different surfaces (`leaves') of the foliation. That is, if the different leaves are parametrised by a real number $s$, then for any pair of proper times $\big ( \tau_r(s), \tau_e(s)\big )$ belonging to the same leaf, the time-dilation effect will be reproduced as in Fig. 3. This is due to the fact that the distribution of positions will be $|\psi(\vec{x}_r, \tau_r(s); \vec{x}_e, \tau_e(s))|^2$ on the leaves of the foliation. But what about a pair of proper times $\big ( \tau_r(s'), \tau_e(s'')\big )$ that belong to different leaves? In this case, the position distribution will \textit{not} \cite{hyper96}, in general, be given by $|\psi(\vec{x}_r, \tau_r(s'); \vec{x}_e, \tau_e(s''))|^2$. The reason for this is simple: the nonlocal velocity field, being foliation-dependent, preserves the $|\psi(\vec{x}_r, \tau_r; \vec{x}_e, \tau_e)|^2$ distribution only on the leaves of the foliation. A different velocity field altogether is required in order to preserve this distribution on the leaves of a different foliation.\\

Nevertheless, a heuristic argument \cite{hyper99} has been given which claims that, for such a model, the disagreement with the $|\psi(\vec{x}_r, \tau_r; \vec{x}_e, \tau_e)|^2$ distribution on hypersurfaces not belonging to the foliation does not entail any experimental violation of the Born rule:
\begin{quote}
``...the predictions of our model are the same as those of orthodox quantum theory... regardless of whether or not these observables refer to a common hypersurface belonging to $\mathcal{F}$ [the preferred foliation]. This is because the outcomes of all quantum measurements can ultimately be reduced to the orientations of instrument pointers, counter readings, or the ink distribution of computer printouts, if necessary brought forward in time to a common hypersurface in $\mathcal{F}$, or even to a single common location, for which agreement is assured.''
\end{quote}
But a heuristic argument cannot -- in principle -- supervene on the mathematical fact that the model predicts the distribution of positions to be different from $|\psi(\vec{x}_r, \tau_r; \vec{x}_e, \tau_e)|^2$ on hypersurfaces not belonging to the preferred foliation. The argument has to be mathematically formulated, including any underlying assumptions, to show that the assertion clearly follows for the scenario considered here. A mathematical approach to this problem has been taken in a recent work \cite{lborn}, which claims to prove that a hypersurface not belonging to the preferred foliation has the correct position distribution when (but not otherwise) detectors are placed over that hypersurface\footnote{Interestingly, placing detectors over a hypersurface then, according to this proof, influences the distribution of particles reaching that hypersurface. This appears to constitute a retrocausal influence of the detectors on the position distribution of particles in the past.}. But since this proof assumes, similar to the previous model, instantaneous collapse across a single preferred frame, applying it to our scenario is problematic. Thus, this model also fails to reach a complete agreement with the predictions made in this article.\\

\section{Discussion}\label{ranna}
We considered a relativistic Bell scenario where one of the wings experiences time-dilation. We generalised the retrocausal Brans model to describe this scenario and found that it predicted a difference in the position distributions between the wings. We also considered the description of our scenario in other hidden variable models. In particular, we argued that $\psi$-ontic models face difficulties in adequately describing the experiment, even if a multi-time quantum state, as in our model, is introduced. \\

The problems faced by the models in section \ref{nanha} are naturally resolved in our model. The essential difference is that the retrocausal Brans model is $\psi$-epistemic; that is, the multi-time quantum state represents the information experimenters have about the joint quantum system. Upon knowing the measurement results, the experimenters update their information about which ontic quantum state, out of the possible ones given $|\psi(\tau_r,\tau_e)\rangle$, their particles belonged to. Thus, instantaneous collapse is not treated as a physical process, but as mere updating of the experimenters' information, which does not require a preferred reference frame for its definition. The problem of defining this frame consistently for our scenario, therefore, evaporates. Secondly, the $\psi$-epistemic viewpoint naturally accommodates, for a `flash' model, the variation of flash probability with respect to different hypersurfaces. The objective situation at a space-time point $(x, t)$, in such a model, is uniquely determined by the relevant ontic variables, but the subjective information about this situation, encoded in the multi-time quantum state, may vary with the hypersurface considered. This variation reflects only the variation in the experimenters' subjective information about the underlying process governing the flashes at $(x, t)$. This resolves the problem of interpreting these probabilities in a relativistic context. Lastly, $\psi(\vec{x}_r, \tau_r; \vec{x}_e, \tau_e)$ does not act as an ontological guiding wave in our model. The velocity field for each particle is, instead, given locally by the ontic quantum states, which remain separable at all times, via equation (\ref{aam}). The multi-time quantum state determines only the distribution of the ontic quantum states for an ensemble, via equation (\ref{multime}). The distribution of positions equals $|\psi(\vec{x}_r, \tau_r; \vec{x}_e, \tau_e)|^2$ for \textit{any} pair of proper times $\big ( \tau_r, \tau_e\big )$. Thus, the problem of reproducing the Born rule distribution for arbitrary hypersurfaces is resolved.\\

Our model, then, arguably suggests that the $\psi$-epistemic interpretation might prove to be the missing ingredient in constructing relativistic generalisations of hidden variable theories of quantum mechanics. For example, there remains, despite several decades of research, a key problem in the program \cite{bohmbook2, hyper96, hyper99, durr14} to build a fundamentally relativistic version of the de Broglie-Bohm theory: any working model requires a preferred foliation of space-time, which violates relativity at the fundamental level. This results from treating the quantum state as an ontological variable, which is problematic from a relativistic viewpoint because the quantum state can be non-separable (entangled), and its associated velocity field, in that case, nonlocal. A $\psi$-epistemic interpretation for this theory is, perhaps, not so surprising when we consider that such a solution will be closer to de Broglie's original idea about his pilot-wave theory being an approximation to a deeper `double-solution' theory \cite{broy, solventini, ecolin}. In his double-solution theory (incompletely worked out), the quantum state has an epistemic status. A $\psi$-epistemic interpretation for objective state reduction models, on the other hand, implies that the probabilities, for spontaneous instantaneous collapses or flashes, are not objective, but subjective. This solves the problem (see the previous section) of physically interpreting the probabilities in a relativistic context, where they turn out to have a dependence \cite{alluri} on the hypersurface considered to evaluate them. It also suggests a deeper mechanism, described by the ontic variables in such a model, underlying the probabilistic description that may turn out to be deterministic. The epistemic interpretation of the quantum state can be inferred from the relativistic concepts of locality and separability in Einstein's preferred version \cite{einstein48, honstein} of the EPR argument \cite{epr}. Taking up this approach may necessarily invoke retrocausality (or superdeterminism), given that such models naturally violate the preparation-independence assumption used in the PBR theorem \cite{pbr} to prove $\psi$-onticity (see the discussion, for example, in ref. \cite{whartonmain}), as well as the measurement-independence assumption used in Bell's theorem to prove nonlocality (see the Introduction).\\

An analysis of the experiment in the standard quantum formalism requires the application of quantum field theory. This is needed to theoretically verify the prediction made by the retrocausal Brans model. However, there are several challenges in using quantum field theory to describe our setup. Firstly, the primary ontological objects in the theory are fields \cite{macha}. Creation and annihilation operators in the theory are concerned with field excitations, and only \textit{heuristically} interpreted in terms of particles. Likewise, an interaction between two fields described by a local interaction Hamiltonian is often heuristically interpreted as a `localised particle interaction'. However, a clear notion of localised particles is required, in our scenario, to ascribe different proper times to the two particles. Defining this notion clearly in quantum field theory has proved a challenge in itself \cite{heger1, heger2, heger3, halo, baran, wall, rovesi, papa} due to a range of issues, from relativistic considerations to the measurement problem. Second, the theory is predominantly used to calculate amplitudes for transitions from an initial state defined at $t \rightarrow -\infty$ to different final states defined at $t \rightarrow +\infty$. On the other hand, the central issue here is not probabilities for various outcomes but the time-evolution of the measurement process. Obtaining the time-dependent description of an interaction between two fields is technically difficult in the theory because the interaction Hamiltonian is time-dependent \cite{macha}. Third, the theory allows for the creation and annihilation of particles, so particle number is not conserved in general \cite{macha}. In addition, the spin operator is not uniquely defined in the relativistic context \cite{chor, caban, friis, palmwest1}. However, this is not a real problem for our setup since the relative velocity of the particle with respect to the Stern-Gerlach apparatus is small at both wings. These challenges have been addressed, mainly in the context of relativistic quantum information theory, to varying degrees. Several authors \cite{cuban1, cuban2, asspin, yonga, hiroshima} have described the EPR and Bell correlations in relativistic settings using quantum field theory, but the time-evolution of the spin measurements, which is the issue at stake here, has not been addressed. The authors of ref. \cite{palmwest1} propose a general formalism to describe localised massive fermions in curved space-time, which they use in ref. \cite{palmwest2} to give a detailed description of relativistic Stern-Gerlach measurements on a single fermion. However, this description of the measurement process is not extended to the case of entangled fermions. We therefore conclude that it is unclear at present, due to the aforementioned difficulties, if quantum field theory reproduces the prediction made by the retrocausal Brans model. If it is, then the retrocausal Brans model can be said to reproduce a prediction of quantum field theory with great simplicity. It would then suggest that a local, separable and $\psi$-epistemic retrocausal hidden variable reformulation of quantum mechanics, if it exists, will have a significant advantage in describing relativistic effects on quantum systems. Such a reformulation will also help us in understanding how to merge the two theories of quantum mechanics and relativity. If the prediction is not reproduced, then one can experimentally test the retrocausal Brans model by an experimental realisation of our scenario. \\

Lastly, we address a few challenges facing the retrocausal approach. From our viewpoint, the most important problem in physically understanding these models is the lack of an explicit retrocausal mechanism. After all, \textit{how} is the information about the future measurement settings made available during the preparation event in the past? The retrocausal Brans model assumes, but does not explain, this transfer of information backwards in time. In the absence of a convincing retrocausal mechanism, one may posit this information to have been made available through a completely different mechanism, for example superdeterministic or even nonlocal, without changing the mathematical structure of the model appreciably. Finding a clear answer to this question will give a physical insight as to how retrocausality works, thereby also reducing the counter-intuitiveness of this approach. A criticism \cite{woodspek} that has received significant attention in recent years \cite{cid1, cid2, cid3, cid4, cid5} is that retrocausal models (in fact, other hidden variable models as well) have to be fine-tuned in order to reproduce the no-signalling nature of Bell correlations. Although this result is framed in the language of causal modelling and fairly recent, the underlying physical argument, that hidden-variable models reproduce the quantum predictions for only very specific initial conditions, was arguably appreciated by several authors (see ref. \cite{bohm54} and the references therein) as far back as the 1950's, in the context of de Broglie-Bohm theory. In the 90's, Valentini \cite{valentinI} gave a natural solution to this problem by explicitly considering non-fine-tuned (``non-equilibrium'') hidden variable distributions and proving that these distributions relax to the fine-tuned (``equilibrium'') distribution, which reproduce quantum predictions, at a coarse grained level due to the dynamics of the theory\footnote{This requires the assumption of no initial fine-grain structure, as in classical statistical mechanics \cite{tallman, davies77} . For an alternative point of view based on typicality of initial conditions, see ref. \cite{92mainstream}. This view has been criticised, in our opinion quite convincingly, by Valentini \cite{teenv}. For a response to the criticism, see ref. \cite{durr19}.}. He clearly discussed the conspiratorial or fine-tuning features of equilibrium, by showing \cite{valentinII, royalvale} how statistical locality emerges in equilibrium and is violated out of equilibrium. He subsequently extended \cite{genon} this discussion to general nonlocal models and proposed that the fine-tuning can be understood by the requirement of a certain `detailed balancing' for nonlocal signals to average to zero. It has subsequently been shown \cite{cv} that the second-order version of de Broglie-Bohm theory does not posssess the necessary relaxation properties. We have argued elsewhere \cite{fpaper} that fine-tuning for retrocausal models may also be approached in a similar manner. That is, given the past and future boundary conditions, the fine-tuning may naturally arise out of the dynamics for some retrocausal models but not for others. The prime benefit of taking such an approach is that it will help us identify the properties of retrocausal models that are important for relaxation to occur. This will rule out several primitve retrocausal models, and help develop future retrocausal models with the appropriate relaxation characteristics. Another approach to the problem, relying on symmetry arguments to rule out the non-fine-tuned distributions, has also been proposed \cite{almada, adlam2}. This approach claims the benefit of using time-symmetry, a central motivation for pursuing retrocausality, to explain the fine-tuning.\\

\section{Funding statement}
This work was supported by the Teaching Assistantship provided by the Department of Physics, Clemson University.

\acknowledgements
I would like to thank Antony Valentini for insightful discussions and helpful suggestions. I am also thankful to Adithya Kandhadai for being a reliable sounding board, Ken Wharton for helpful comments on an earlier draft of the paper, and to an anonymous referee for pointing out several interesting multi-time models.

\bibliographystyle{bhak}
\bibliography{bib}

\end{document}